**TITLE: Analysis of the Institutional Free Market in Accredited Medical Physics Residency Programs**

Alexander Niver[1] & Brian W. Pogue[1,2*]

[1]Department of Medical Physics, University of Wisconsin School of Medicine and Public Health, Madison WI 53705

[2]Thayer School of Engineering at Dartmouth, Hanover NH 03755

*Contact author:  bpogue@wisc.edu or brian.w.pogue@dartmouth.edu

**ABSTRACT:**

***Purpose:*** The pathway to clinical medical physics involves 3 distinct stages of accredited positions, including (i) graduate education, (ii) residency training & (iii) employment. While CAMPEP accredits programs & people, the type of positions and numbers available are each more influenced by local institutional free market forces driving their choices.

***Methods:*** Understanding the overall status and trajectories of the full range of residency positions was the rationale for this survey, using publicly available data reported 2017 to 2023 timeframe.

***Results:*** Current data shows 108 therapy residency programs with 163 positions, and 38 diagnostic imaging residency programs with 39 positions. The total numbers show growth in positions by +6.5/year in therapy, and +2.1/year in imaging. There are 18 therapy and 10 imaging residencies longer than 2 years, all at university hospitals, and nearly all requiring a PhD.  The number of programs that accept graduates with either MS or PhD, versus PhD only, has grown, from 69 to 105 from 2017 to 2023. There was a trend in large health networks increasing their residency numbers from 9 to 23 in this timeframe. The total 7-year growth trend was compared to numbers of graduating students, showing reasonable agreement between  positions with expected inter-year variation.

***Conclusions:*** There has been growth in numbers of accredited residencies, with accompanying shifts in program types & student education expected. Most are 2-years, while there is smaller growth in longer programs with research added in, primarily at university hospitals. Major growth has occurred in residencies accepting both MS and PhD graduates, while those accepting only PhD graduates tends to be more imaging focused and at university centers. Overall, the number positions has been keeping pace with the known growth in graduates from accredited medical physics graduate programs.

**1.0 Introduction**

Education and training in the field of clinical medical physics is carried out by individual institutions who have CAMPEP accreditation. While the accreditation process is regulated by CAMPEP [1], the number of such opportunities is nearly entirely driven by the goals and financial mechanisms unique to each individual institution. The residency experience is a unique one which is between graduate education and before employment as a physicist, and yet there are few direct organizational links between them. For example, the number of any of trainees at either the graduate students admitted or graduating are not coordinated or linked to the number of resident opportunities, and these are not directly linked to available positions [2]. Rather the number of each is more linked to financial decisions of what is possible at each individual institution. As such, there is some value in understanding the program variations, numbers of students at different training levels, and if these match the opportunities and needs of the field. In this study an overview of the available residency positions was carried out using publicly available data.

Residency training is a critical component of educating new medical physicists, which is why the accreditation standards are well thought out and key characteristics are prescribed to the institutions [3, 4]. This accreditation is largely about the experience of the mentors and trainees in the program, rather than any aspect of admission numbers or support processes. Aspects, such as number of slots admitted, training background desired, number of years in the program, and existence of research are all part of individual program design. Programmatic independence between institutions and recent changes to heathcare practice systems have allowed for a growing variety in residency programs, resulting in some shifting trends in recruitment of new residents over time. This current study analyzes these temporal trends, and compares them to the results from a previous study [5] regarding the education of medical physics graduate students at CAMPEP accredited programs. The data is organized to help those in the education and employment sectors of medical physics understand where the residency opportunities are as an overview snapshot today.

**2.0 Methods**

All of the survey data came from the requirement that all CAMPEP residencies self-publish their own data on their own program's website. In some cases where this was not updated locally, an email follow-up was done with individual program directors, as listed below. These data included the number of open positions for new residents for each year in the study (2017-2023), the minimum level of education required to apply, duration of residency, and the type of institution hosting the residency. Institutions were classified as private companies (e.g. Varian), health care networks (e.g. Texas Oncology) or university/university hospitals (e.g. Duke University Medical Center).

All websites were accessed for this data between January and March of 2025, and any needed follow up with individual program directors was in this same timeframe. Of the 141 CAMPEP accredited residencies surveyed, 105 had published current data on their websites. For the remaining 36, the residency directors were directly emailed in an attempt to obtain the missing information, of which 20 responded, leaving 16 programs with incomplete data sets. Regarding the data of most interest, resident positions available in 2017 and 2022-2023, only 4 programs were missing these specific data, suggesting that there was less than approximately 2-3% error in this survey.

The study separated the data into two primary groups: therapy physics residencies and diagnostic imaging physics residencies. Over the length of the study period 31 therapy physics residencies received their initial CAMPEP accreditation as did 24 diagnostic imaging physics residencies. These new residencies (recently accredited) were included in the overall trends but were also separately analyzed. Previously established residencies, those which received their accreditation in 2016 or earlier or were operating as medical physics residencies before receiving CAMPEP accreditation, were also analyzed to see whether changes in the education of residents were primarily from changes in new or established residencies. Nuclear medicine physics was not included in this study and residencies which closed within the study period were also not included.

This study also reviewed work done in a previous publication [5] to analyze how trends in residency training compare to graduations from CAMPEP accredited medical physics graduate programs.

## 3.0 Results

### *3.1 Growth in Accredited Medical Physics Residencies*

Between 2017 and 2023, 24 therapy physics and 18 diagnostic imaging physics residencies received their accreditations. This means that 21.2% of therapy physics and 63.1% of diagnostic imaging physics residencies were newly accredited within this surveyed 2017-2023 period. **Figure 1a** shows the number of accredited residencies on a year-by-year basis, separating imaging and therapy and also showing the cumulative sum of them. There was clear growth in the number of resident positions during this time, as shown in **Figure 1b**. In 2017 there were 114 therapy and 24 imaging positions available, which grew to 138 therapy and 34 imaging by 2023. This growth is shown in detail in figure 1b, with average annual growth numbers of 6.5/year in therapy and 1.4/year in imaging.

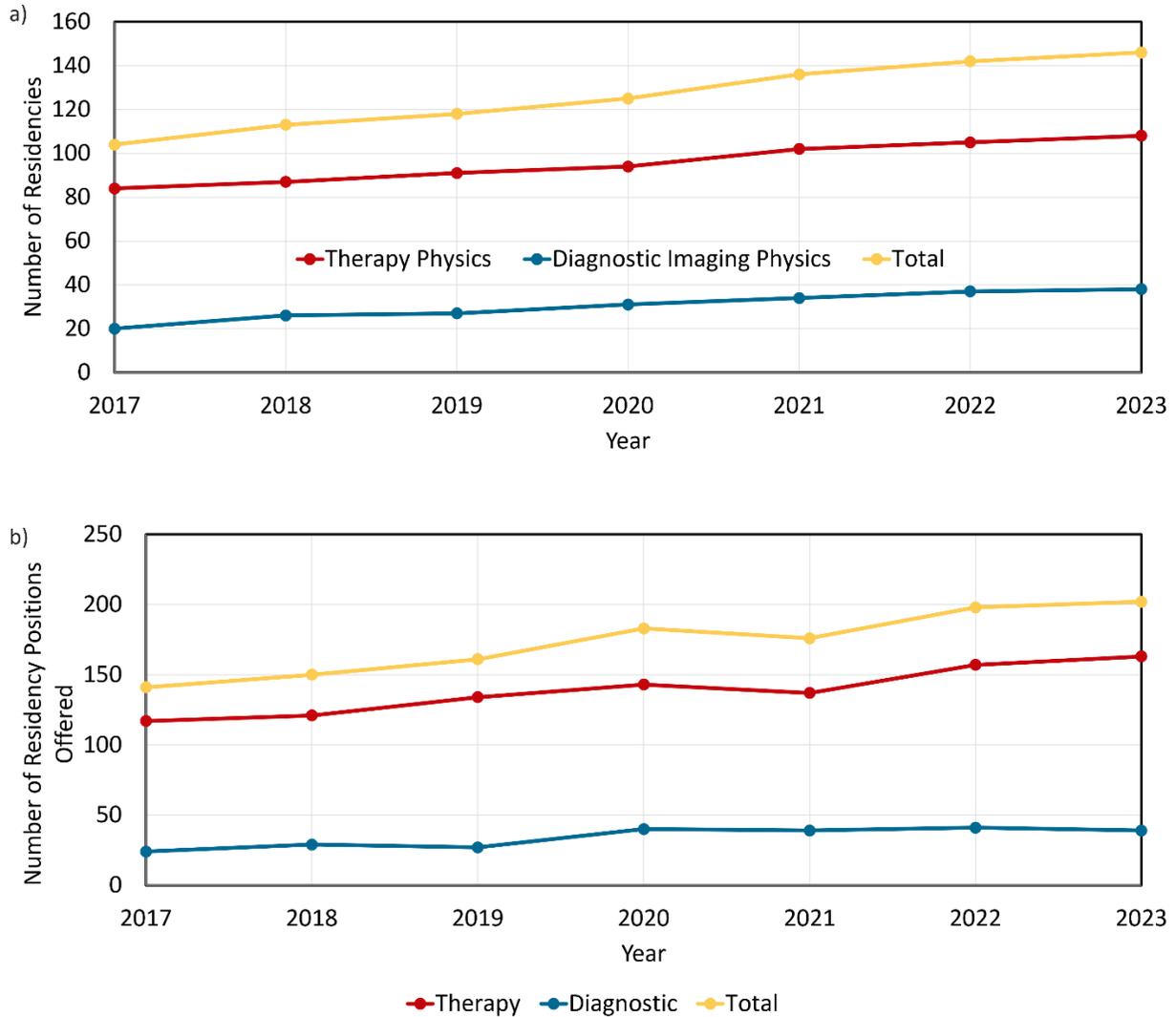

**Figure 1.** Survey data accumulated from all CAMPEP websites, showing 6 year trends in numbers of residency institution programs (a) and total residency positions (b).

When separated into those programs which were accredited during the study period and those that were previously established, the growth in positions available to new residents in therapy physics residencies grew roughly evenly between the two sub-groups, but with a greater effect from the previously established programs. Over the study period there was an average increase of 3.1 new positions/year from recently accredited and 3.4 new positions/year from previously established programs. This was not the case for diagnostic imaging physics residencies which saw growth primarily from the recently accredited programs with 1.7 new positions/year while the number of positions available to incoming residents remained largely unchanged in previously established programs, growing only 0.43/year new positions.

When comparing these numbers to CAMPEP graduate education data, from a previous study over the same study period, there were more positions for new residents than graduates. Starting in 2021 there were fewer positions for new residents than CAMPEP graduates. There was a significant increase in CAMPEP graduates in 2022 outpacing the residency positions for that year. The differential shortage was reduced by 2023, and all these trends are shown in **Figure 2**.

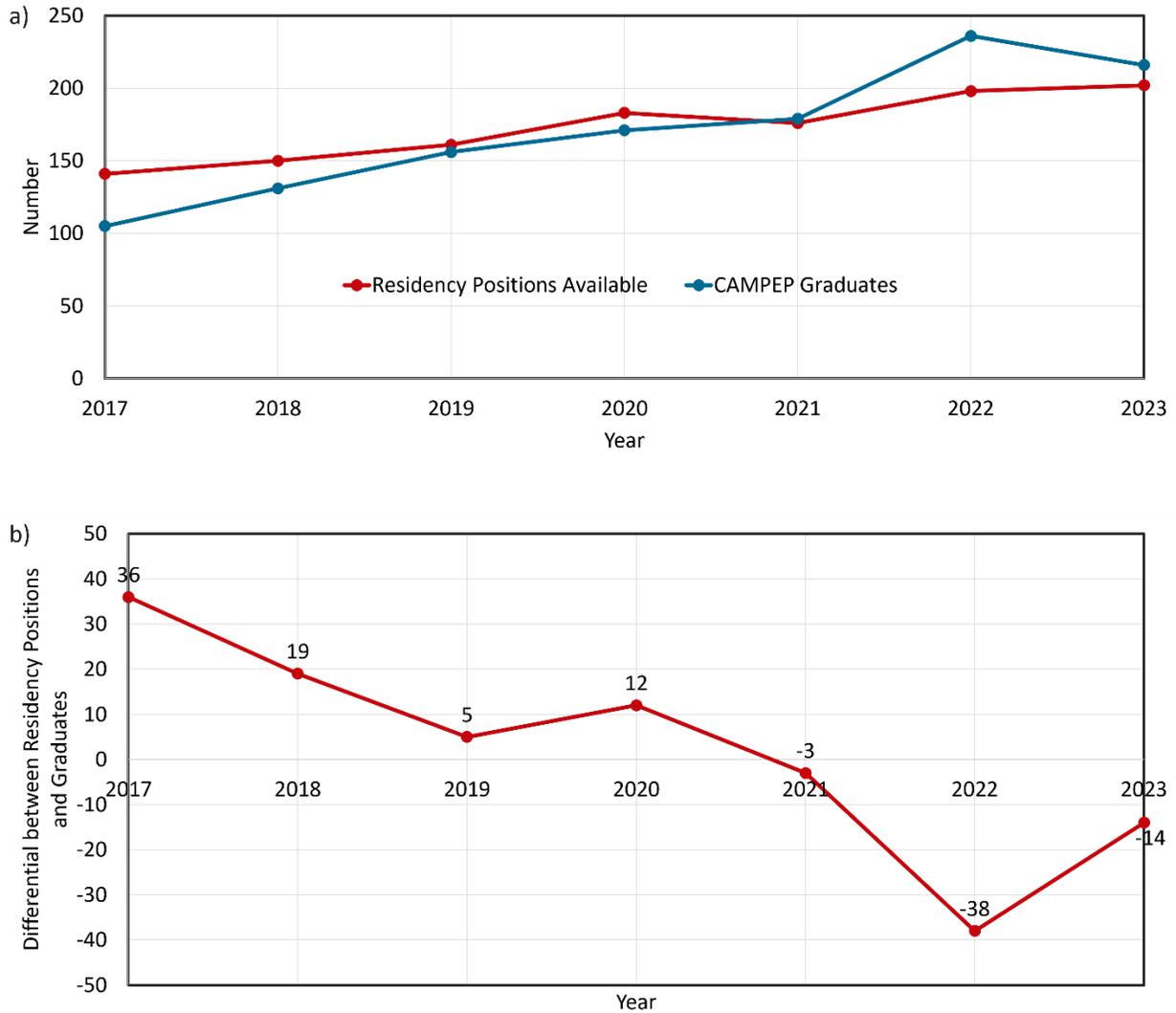

**Figure 2.** Total annual residency positions are listed for the survey period (a) with addition of listings of graduate numbers from MS and PhD programs [5]. In (b) the annual differential between graduates and residency positions is shown.

### *3.2 Residency Program Length & Institution Type*

The vast majority of residency programs, for both therapy physics (82.4%, 84 of 102) and diagnostic imaging physics (76.9%, 30 of 39), train their residents on a 2-year cycle. A small but significant portion of residencies offer an additional, third, year for residents to perform research. This is more prevalent among diagnostic imaging physics residencies (20.5%, 8 of 39) than therapy physics residencies (10.8%, 11 of 102), shown in **Figures 3 (a) and (b).** Another option for residents is variable length residencies where a program will offer a 2-year residency for a certain number of residents and longer terms for others (e.g. 1 resident would be hired for 2 years and another for 3 years). Such options are rare, represented by less than 6% of residencies.

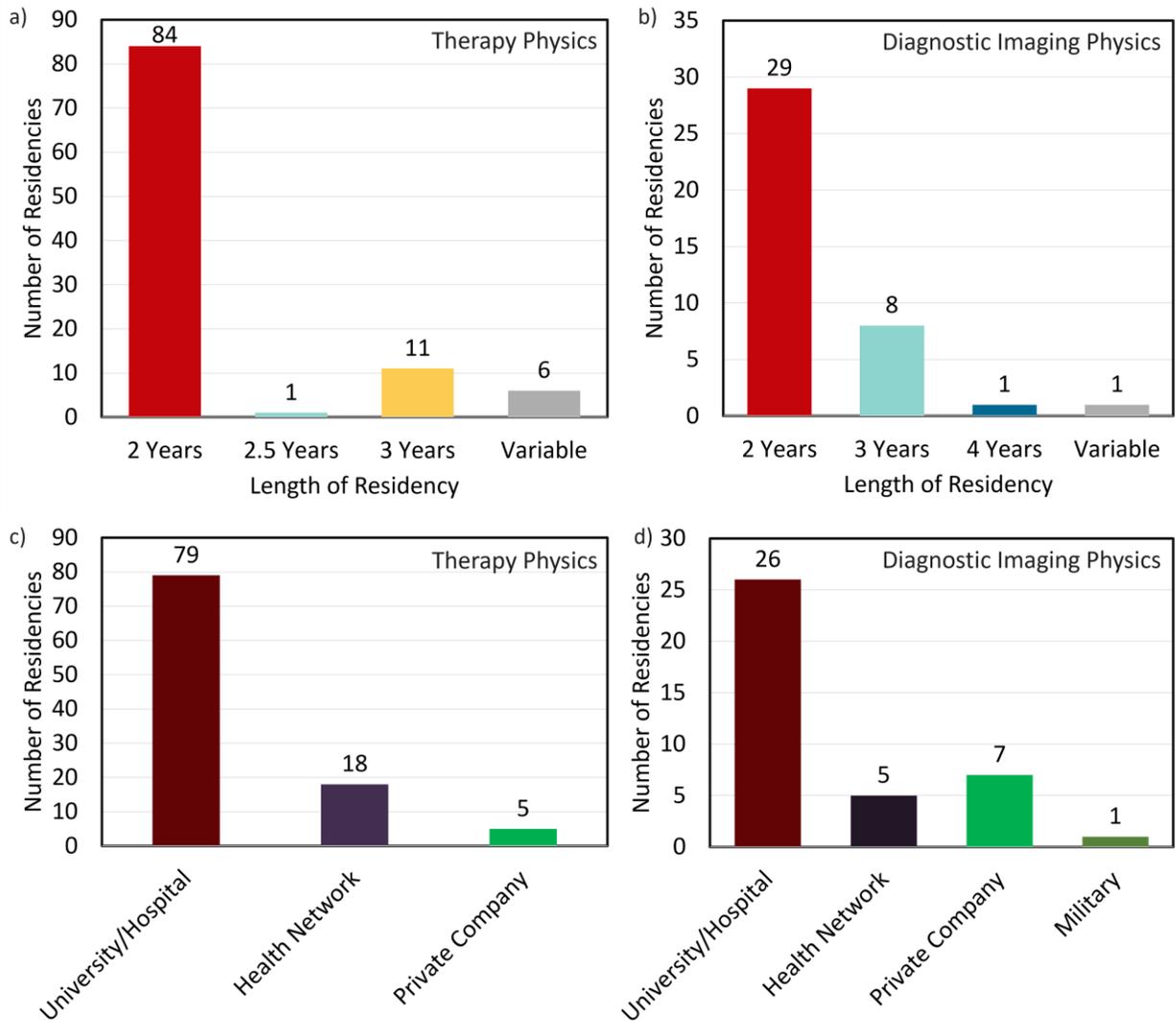

**Figure 3.** Residency numbers are listed with years of the program for (a) therapy and (b) diagnostic programs. The types of institutions that offer the residency are listed in (c) and (d) for therapy and diagnostic programs.

This study then investigated what kind of organizations host medical physics residencies. They were divided into 4 groups: for-profit private companies, health care networks, universities/university hospitals and military, shown in **Figure 3 (c) and (d)**. The majority of all medical physics residencies are hosted at universities/university hospitals with 77.5% of therapy physics residencies (79 of 102) and 66.7% of diagnostic imaging residencies (26 of 39). A full illustration of how many residencies are hosted by each organization type is provided in the figure.

### *3.3 Graduate Training Expectation: MS & PhD*

A final aspect examined in the data was the minimum education requirements for application to each residency program. The programs are required to list the education requirements they are looking for in candidates as either PhD only or open to MS and PhD graduates. There is a roughly equal fraction (23%) in both of these for therapy physics (23 of 102, 22.5%) and diagnostic imaging physics (9 of 39, 23.1%). The data is shown in figure 4(a) and (b) for 2023, as a trend over the study period in (c). The number open to MS students has increased from 71% (69 of 96 in 2017) to 77% (105 or 136 in 2023), showing a steady but clear opening up in acceptances for the broader range of training. All of the diagnostic imaging physics residencies that require a PhD fall under the university/university hospital category and 22 of the therapy physics residencies do so as well. Programs hosted at universities/university hospitals also make up most residencies longer than two years with 16 of the 18 longer-term residencies for therapy physics and 8 of 10 for diagnostic imaging physics being at universities/university hospitals.

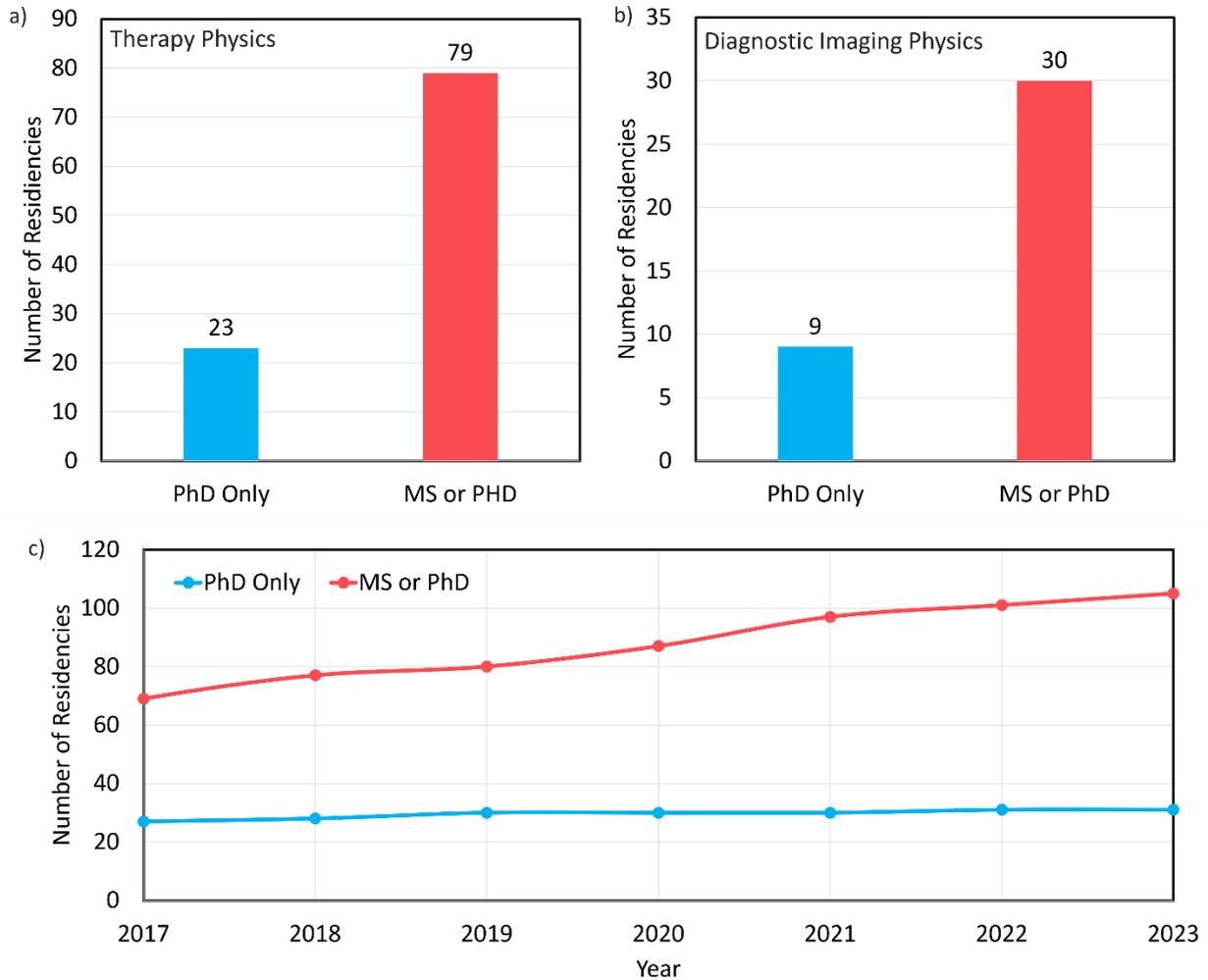

**Figure 4.** The admission goals for the residency programs are divided in terms of seeking either PhD only candidates, or open to MS or PhD graduates, and are listed for (a) Therapy programs, (b) diagnostic imaging programs (b) and as a trend over time for all programs (c).

## 4.0 Discussion

### 4.1 Growth Trends in Residencies

There is growth in both the number of CAMPEP accredited residency programs and in the number of positions available to new residents per year for both therapy physics and diagnostic imaging physics programs. The growth of new resident positions in diagnostic imaging physics residencies from the newly established programs also corresponds with the fact that over 60% of all diagnostic imaging physics residencies were accredited within the study window and contributes to the fact that most of the growth in diagnostic imaging physics resident numbers has come from these newly established programs.

While the data is not 100%, the margin of error in this survey is thought to be very small.  Of the 102 therapy physics residencies, it was not possible to obtain information for the year 2023 from just 3 programs, all of which fall under the category of newly established programs. This did not significantly affect the data as the 3 missing programs each recruited 1 resident per year over the other years of the study for which we have data.

### 4.2 Comparison to CAMPEP Graduating Student Data

Over the study period the difference between the numbers of positions available to new residents and CAMPEP graduates in the same year have fluctuated and presented two significant peaks shown in figure 2b: the first in 2017 with 36 more available resident positions than CAMPEP graduates and the second in 2022 with 38 more CAMPEP graduates than available resident positions. In 2017, 13 residencies obtained their accreditation which added 10 new resident positions and 2017 was also prior to the significant increase in MS graduates from CAMPEP accredited graduate programs which began in 2018. The peak in 2022 corresponds with a spike in the number of graduates, both MS and PhD, which was partially mitigated by an increase in residency positions offered.

### 4.3 Residency Institutions

As shown in Figure 3, most medical physics residencies are hosted at universities/university hospitals, which are institutions more likely to be associated with education and training as part of their mandates, as well as being suitably equipped and staffed to meet the requirements for a residency as laid out in AAPM report 90. They may also more easily meet the minimum requirements to have either 500 patients undergoing radiation therapy or 50,000 diagnostic imaging procedures per year. When comparing the length of a residency with where it is hosted, the vast majority of residencies (24 of 28) that last longer than 2 years are hosted at universities/university hospitals. Presumably this is because these longer residencies offer opportunities for residents to perform additional research, where university hospitals have an explicit mandate to advance this area. University or university hospital based residencies also account for almost all programs that require a PhD to apply, though there is no correlation between the length of the residency and the requirement of the applicant to have a PhD.

Health networks comprising multiple hospitals and/or clinics have been a growing portion of medical physics residences, as consolidate into networks has been a growing national trend in all of healthcare in the past few decades. At the beginning of 2017 there were 9 health network-based programs which made up 10.6% of all medical physics residencies in the US, and in 2023 they represented 15.5% (18/102 therapy physics and 5/39 diagnostic imaging physics) of residencies in the US. During the study window 14 health network based residencies were accredited which accounts for over half (60.8%) of all such residency programs.  Total data on this is shown in Figure 3b.

Residencies hosted by private companies are more common for diagnostic imaging physics than for therapy physics, both as a fraction of their individual totals (17.9% for diagnostic imaging physics vs. 4.9% for therapy physics) and as raw numbers (7 for diagnostic imaging physics vs 5 for therapy physics). There are several companies that provide medical physics consulting services such as shielding design, hardware accreditation, and acceptance testing and when these companies host a medical physics residency, they favor diagnostic imaging physics. There are also companies that provide locum physics services and focus on acceptance testing and hardware accreditation for radiation therapy, which when they host a residency will instead focus on therapy physics. Another important type of private company based residency is those hosted by companies that manufacture medical physics hardware and testing equipment. While few, these residencies offer positions to several residents. The Varian Advanced Oncology Solutions residency offered an average of 4.1 residency positions for new residents every year over the course of the study window. Landauer is also, at the time of writing, currently seeking accreditation for their diagnostic imaging physics residency which would offer 2 positions to new residents per year. This is significantly more than the average across diagnostic imaging physics which was 0.95 positions for new residents per residency in 2023.

### 4.4 Residency Education Requirements

A temporal analysis of the requirement for an incoming resident to have a PhD shows that this requirement has remained at similar numbers in recent years. During the study window, only 5 programs were accredited that required a PhD, and from 2020-2023 only 1 new program was accredited with this requirement, while 48 new programs were accredited overall that offered positions to graduates with either an MS or PhD degree. Figure 4 shows this trend in greater detail.

### 5.0 Conclusion

The number of CAMPEP accredited medical physics residencies in the US has grown significantly over the course of the study period, as has the number of positions offered to new residents every year. Growth in the number of residencies was more pronounced in the field of diagnostic imaging physics as roughly half of such residencies were accredited during the study period while growth in the number of positions for new residents has been more significant in therapy physics residencies. Both groups of medical physics residencies favor 2-year programs, but those hosted by universities and university hospitals tend to offer longer-term programs with opportunities for research. Residency programs hosted by health networks and private companies tend to favor 2-year programs and are also growing in prevalence. The largest trend overall is a growth in number of residencies that accept both MS and PhD graduates, while those accepting only PhD graduates tends to be more imaging focused and more based at university medical centers. Overall, the number of offerings of residency positions for new residents has been keeping pace with the known growth in the number of graduates from CAMPEP accredited medical physics graduate programs year-on-year.

### Conflict of Interest

The authors have no conflicts of interest relevant to the content of this paper.

**References**

1. *CAMPEP Accredited Graduate Programs in Medical Physics*. Available from: http://campep.org/campeplstgrad.asp.
2. Bayouth, J.E., J.W. Burmeister, and C.G. Orton, *Point/counterpoint. Medical physics graduate programs should adjust enrollment to achieve equilibrium between graduates and residents*. Med Phys, 2011. **38**(8): p. ii-iv.
3. Prisciandaro, J.I., C.E. Willis, J.W. Burmeister, G.D. Clarke, R.K. Das, J. Esthappan, B.J. Gerbi, B.A. Harkness, J.A. Patton, D.J. Peck, R.J. Pizzutiello Jr., G.A. Sandison, S.L. White, B.D. Wichman, G.S. Ibbott, and S. Both, *Report No. 249 - Essentials and Guidelines for Clinical Medical Physics Residency Training Programs (2013)*. 2013.
4. Lane, R.G., D.M. Stevens, J.P. Gibbons, L.J. Verhey, K.R. Hogstrom, E.L. Chaney, M.C. Martin, E.E. Klein, K.P. Doppke, B.R. Paliwal, R.E. Wendt III, and M.G. Herman, *Report No. 090 - Essentials and Guidelines for Hospital-Based Medical Physics Residency Training Programs (2006)*. 2006.
5. Niver, A. and B.W. Pogue, *Analysis of the Institutional Free Market in Accredited Medical Physics Graduate Programs*. J Appl Clin Med Phys, 26(7) e70164, 14 July 2025 https://doi.org/10.1002/acm2.70164